\documentclass[a4paper]{article}

\usepackage[sort]{cite}
\usepackage{listings}
\usepackage{booktabs}

\usepackage{url}
\urldef{\mymail}\path|tresoldi@gmail.com|

\begin{document}

\title{Newer method of string comparison: the Modified Moving Contracting Window Pattern Algorithm}
\author{
Tiago Tresoldi \\
Universidade Federal do Rio Grande (FURG) \\
\mymail
}
\maketitle

\begin{abstract}
This paper presents a new algorithm, the Modified Moving Contracting Window
Pattern Algorithm (CMCWPM), for the calculation of field
similarity. It strongly relies on previous work by Yang \textit{et al.} (2001),
correcting previous work in which characters marked as inaccessible
for further pattern matching were not treated as boundaries between
subfields, occasionally leading to higher than expected scores
of field similarity. A reference
Python implementation is provided. 
\end{abstract}

\section{Introduction}

Assessing string similarity (or, as commonly referred in the literature, ``field
similarity''), is a recurrent problem in computer science, particularly when dealing with
natural language and genetic data. 
A review of relevant literature on the topic
is presented by Yang \textit{et al.} \cite{YANG01}, which
stresses the importance
of these methods in applications such as data searching and cleansing, Web searching,
computational biology, and data compression.

These areas of research were considered by the authors
before introducing their
method for field comparison, the Moving Contracting Window Pattern Algorithm
(MCWPA). MCWPA is fundamentally different from the general
methods for field comparison based in
measures of ``edit distance'' (such as the
one proposed by Wagner \textit{et} Fischer, \cite{WF74}), which expand
initial work by Vladimir Levenshtein \cite{Lev66}. These methods,
commonly referred to under the single label of ``Levenshtein Distance'',
are generally defined as counts of the minimum number
of single character ``edits'' required to mutate a field $F_{X}$ into a field $F_{Y}$,
where ``edit'' is defined as either an insertion, a deletion, or a substitution
of a single character;
in general, no difference in weight for the various types of ``edit'' is specified.
The method for this computation of field similarity
is closely related to pairwise field alignment and usually
implemented after the aforementioned Wagner-Fischer algorithm or through
dynamic programming approaches such as the one proposed by Guseld \cite{GUS97}.

The algorithm proposed by Yang \textit{et al.} 
extends and generalizes an alternative ``token--based'' approach by 
Lee \textit{et al.} \cite{LL+99},
developed in the context of data cleansing. In their paper, the authors present a pseudo-code
for the algorithm, claiming that their solution
``not only achieve[s] higher accuracy but also gain[s] the time
complexity O(knm) (k $<$ 0.75) for worst case'',
comparing the accuracy of their proposal
with the one of the method by Lee \textit{et al.}
and concluding that
``[t]heoretical analysis, concrete examples and experimental result show that
[the proposed] algorithms can significantly improve the accuracy and time complexity of the
calculation of Field Similarity''.

In the course of a research conducted around 2005 with extensive usage of
the Natural Language Toolkit (NLTK), a Python library and framework for Natural Language
Processing by Bird \textit{et al.} \cite{NLTK},
the author of the current paper needed to perform hundreds of field comparisons
for sorting lists of fields according to 
their similarity to a number of
reference field, usually short strings containing natural language data.
Levenshtein distance, the most recommended method, proved slow
when computed without dynamic methods and, more importantly, was found to
be unsuitable for a considerable number of cases, as its scores of similarity,
adjusted to ratios between $0.0$ and $1.0$,
failed to match the magnitude of similarity that most speakers of the natural languages
in study would expect or report. A theoretical investigation suggested that
the obstacle was
an intrinsic limitation of the algorithm itself
given by its focus in general field comparison (i.e., with no \textit{a priori}
assumptions on the entropy of both fields), and was triggered by
idiosyncrasies of the morphology and the orthography of the languages in analysis.
While the difficulties could in part be circumvented with a combination of orthographic,
phonological and, exceptionally, morphological mappings, 
the decision rested in adopting new methods.
A bibliographic research suggested the paper by Yang \textit{et al.}, and,
while for our purposes 
the new algorithm performed better than the edit distance method,
its results were occasionally unexpected and, eventually, worse than 
Levenshtein distance for
some corner cases. The author wrote a revised version that partially solved
the deviations, and the 
Python module which implemented them was eventually
included among the ``contributions'' to NLTK\footnote{The module has
apparently been removed from newer versions of NLTK, but can be easily
found in public forks based on older versions of the toolkit.}.

When needing to perform a similar task almost a decade after that first revision,
the author decided to write a new version which fully and correctly implemented
the revised method, presenting the algorithm and its implementation in this paper.

\section{Background}

A brief but throughly description of the ``token-based'' approach for the computation of
field similarity proposed by Lee \textit{et al.} is given in the second
section of Yang \textit{et al.}, from which the outline of the current section is developed.

Let a field $X$ of length $n$ be composed of tokens (such as ``words'') 
$T_{X_{1}}$, $T_{X_{2}}$, $\ldots{}$, $T_{X_{n}}$ and the corresponding field $Y$ 
of length $m$ be composed of tokens
$T_{Y_{1}}$, $T_{Y_{2}}$, $\ldots{}$, $T_{Y_{m}}$. Each token $T_{X_{i}}$,
where $1 \leq i \leq n$, is compared with each token $T_{Y_{j}}$,
where $1 \leq j \leq m$. Let $DoS_{X_{1}}$, $DoS_{X_{2}}$, $\ldots$, $DoS_{X_{n}}$,
$DoS_{Y_{1}}$, $DoS_{Y_{2}}$, $\ldots$, $DoS_{Y_{m}}$ be the maximum degree of similarity
for tokens $O_{X_{1}}$, $O_{X_{2}}$, $\ldots{}$, $O_{X_{n}}$,
$O_{Y_{1}}$, $O_{Y_{2}}$, $\ldots{}$, $O_{Y_{m}}$, respectively.
The Field Similarity between $F_{X}$ and $F_{Y}$ is computed as follows:

\begin{equation}
SIM_{F_{(X, Y)}}=\frac{\sum_{i=1}^{n}DoS_{X_{i}} + \sum_{j=1}^{m}DoS_{Y_{j}}}{n+m}
\end{equation}

The algorithm proposed by Yang \textit{et al.} generalizes the ``tokens'' employed by
Lee \textit{et al.}, essentially
words in natural languages, into ``window patterns'', which are defined as subfields of minimal
length equal to $1$. As in the first example given in their paper, for the string \texttt{"abcde"},
considering a window of size $3$ sliding from left to right, the series of patterns obtained
is composed of \texttt{"abc"}, \texttt{"bcd"}, and \texttt{"cde"}.
The field similarity in MCWPA is given by the sum of the squares of the
number of the same characters, or minimal units, 
between fields $F_{X}$ and $F_{Y}$, which is defined as the cumulative sum of the
square of combined length of minimal units matched in both fields, i.e. twice the
length of the pattern; the sum is accumulated
while marking already matched subfields as inaccessible for further comparisons.

Thus, in MCWPA,
let a field $F_{X}$ of $n$ characters and a field $F_{Y}$ of $m$ characters; the field
similarity between the two fields, 
which ``approximately reflects the ratio of the total number of the common characters in
two fields to the total number of characters in two fields'',
where SSNC represents the Sum of the Square of the
Number of Same Characters between $F_{X}$ and $F_{Y}$, is computed as follows:

\begin{equation}
SIM_{F_{(X, Y)}} = \sqrt{\frac{SSNC}{(n+m)^{2}}}
\end{equation}

The algorithm is described in depth by Yang \textit{et al.}, with a number of examples 
and graphical
representations of the inner workings of the sliding window approach. 

\section{Changes to MCWPA}

The author of the current paper first implemented the MCWPA algorithm in Python following
the pseudo-code given in Figure 1 in Yang \textit{et al.}. While the authors did not offer
actual code or reference values to test implementations of theirs algorithms,
all the examples could be matched, suggesting that the implementation was correct.

When testing the implementation in production code, however, it was verified that for some
corner cases the results returned were unsuitable, 
with scores generally higher than what was expected by human reviewers.
The author also experimented with some
random strings used in imitation of genetic data, generated by an \textit{ad hoc}
weighted random function according to a table of DNA codon frequencies for the
human genome; for a restricted number of test samples the results
were considered equally unsatisfactory.

An investigation of the algorithm did not prove sufficient for identifying
any theoretical limit as a source for the unsatisfactory scores.
After implementing the
algorithm in multiple and different ways, by a trial and error methodology
an hypothesis was developed that the problem resided in a simplification of
the original implementation, which can be found in the pseudo-code itself
and might have been intentional, as MCWPA is less computationally expansive
than the revised method here proposed and the limitation 
affected a small number of cases.

In detail,
while the theoretical description of the paper and the pseudo-code correctly call
for marking characters in $F_{X}$ and $F_{Y}$ as ``inaccessible'' after a 
given pattern matching, 
the implementation was apparently not marking 
inaccessible characters as a boundary for future pattern matchings, thus allowing
new windows to ``jump over them''. The limitation might
have been introduced when adapting the method from a token based to
a character based approach, as the implementation in Lee \textit{et al.} doesn't
seem to allow non contiguous tokens to be matched. This hypothesis
cannot be verified without access to the original source code.

To illustrate the difference of the algorithm here proposed, the implementation
yielding results considered wrong would,
when matching the strings \texttt{"A123B"} and \texttt{"123AB"},
first match the pattern of characters \texttt{"123"}, but
after the deletion of this pattern in both fields it would not treat the
residual characters as groups of non-overlapping and non-contiguous
subfields (\texttt{"A"} and \texttt{"B"} for the first field,
\texttt{""} and \texttt{"AB"} for the second),
but as two identical \texttt{"AB"} strings.
When reducing the length of the window from $3$ to $2$, the implementation
would incorrectly find a match of a substring of length $2$, when, from the theoretical
stand point of the algorithm, it would be supposed to identify two 
different matches of length $1$, with a lower final score.

The Python implementation presented in the following section solves this
problem by replacing the operation of string concatenation 
of the first version by operations on lists, introducing the
concept of ``sub-fields'', i.e., non-overlapping and non-contiguous factors resulting
after the removal of a specific factor (the pattern being matched) from
a starting field or subfields. When the algorithm matches a pattern, it
returns two subfields, i.e., the characters that precede and the
characters that follow the matched pattern, if any.
As the subfields might be empty, when the match includes the first or
the last character in the string, a check is performed to filter out
such empty subfields from the list of subfields.

As stated, the distortion was only found in corner cases, and actual
scores were higher than the expect only to a reduced limit.
However, as
bibliographical research did not find any mention or correction to Yang \textit{et al.},
even though their paper has a considerable number of citations,
the author found it useful to publish
this corrected implementation under the name of MMCWPA
(Modified Moving Contracting Window Pattern Algorithm). The
author wishes to publicity note that his implementation distributed with old versions
of NLTK at the time of writing is still affected by the problem describe above,
and should be replaced whenever possible by the one presented here.

\section{Python implementation}

A stand-alone Python implementation for the algorithm is presented in this
section. As per
the terms of the ``MIT License'', permission is hereby granted, free of charge, to any
person obtaining a copy of this software and associated documentation files
(the ``Software''), to deal in the Software without restriction, including without
limitation the rights to use, copy, modify, merge, publish, distribute, sublicense,
and/or sell copies of the Software, and to permit persons to whom the Software is
furnished to do so, subject to the following conditions:

\begin{itemize}
\item The above copyright notice and this permission notice shall be included in all copies
or substantial portions of the Software.

\item The software is provided ``as is'', without warranty of any kind, express or implied,
including but not limited to the warranties of merchantability, fitness for a
particular purpose and noninfringement. In no event shall the author
be liable for any claim, damages or other liability, whether in an action of
contract, tort or otherwise, arising from, out of or in connection with the software
or the use or other dealings in the software.
\end{itemize}

\begin{lstlisting}[language=Python,showstringspaces=false,frame=single,caption={CMCWPA}]
def mmcwpa(f_x, f_y, ssnc):
  """
  An implementation of the Modified Moving Contracting
  Window Pattern Algorithm (MMCWPA) to calculate string
  similarity, returns a list of non-overlapping,
  non-contiguous fields Fx, a list of non-overlapping,
  non-contiguous fields Fy, and a number indicating the
  Sum of the Square of the Number of the same
  Characters. This function is intended to be "private",
  called from the "public" stringcomp() function below.

  @param f_x: A C{list} of C{strings}.
  @param f_y: A C{list} of C{strings}.
  @param ssnc: A C{float}.

  @return: A C{list} of C{strings} with
           non-overlapping, non-contiguous subfields
           for Fx, a C{list} of C{strings} with
           non-overlapping, non-contiguous subfields
           for Fy, and a C{float} with the value of
           the SSNC collected so far.
  @rtype: C{list} of C{strings}, C{list} of C{strings},
          C{float}
  """

  # the boolean value indicating if a total or partial
  # match was found between subfields of Fx and Fy; when
  # a match is found, the variable is used to cascade
  # out of the loops of the function
  match = False

  # the variables where to store the new collections of
  # subfields, if any match is found; if these values
  # are not changed and the empty lists are returned,
  # stringcomp() will break the loop of comparison,
  # calculate the similarity ratio and return its value
  new_f_x, new_f_y = [], []

  # seach patterns in all subfields of Fx; the index of
  # the subfield in the list is used for upgrading the
  # list, if a pattern is found
  for idx_x, sf_x in enumerate(f_x):
    # `length` stores the length of the sliding window,
    # from full length to a single character
    for length in range(len(sf_x), 0, -1):
      # `i` stores the starting index of the sliding
      # window in Fx
      for i in range(len(sf_x)-length+1):
        # extract the pattern for matching
        pattern = sf_x[i:i+length]

        # look for the pattern in Fy
        for idx_y, sf_y in enumerate(f_y):
          # `j` stores the starting index in Fy; the
          # Python find() function returns -1 if there
          # is no match
          j = sf_y.find(pattern)
          if j > -1:
            # the pattern was found; set `new_fx` and
            # `new_fy` to version of `fx` and `fy` with
            # the patterns removed, update the SSNC and
            # set `match` as True, in order to cascade
            # out of the loops

            tmp_x = [sf_x[:i], sf_x[i+length:]]
            tmp_y = [sf_y[:j], sf_y[j+length:]]
            new_f_x = f_x[:idx_x] + tmp_x + f_x[idx_x+1:]
            new_f_y = f_y[:idx_y] + tmp_y + f_y[idx_y+1:]

            ssnc += (2*length)**2

            match = True

            break

          # if the current pattern was found, end search
          if match:
            break

        # if a match was found, end sliding window
        if match:
          break

      # if a match was found, end Fx subfield enumeration
      if match:
        break

  # remove any empty subfields due to pattern removal
  new_f_x = [sf for sf in new_f_x if sf]
  new_f_y = [sf for sf in new_f_y if sf]

  return new_f_x, new_f_y, ssnc

def stringcomp(str_x, str_y):
  len_x, len_y = len(str_x), len(str_y)

  f_x, f_y = [str_x], [str_y]

  ssnc = 0.0
  while True:
    f_x, f_y, ssnc = mmcwpa(f_x, f_y, ssnc)

    if len(f_x) == 0 or len(f_y) == 0:
      break

  return (ssnc/((len_x+len_y)**2.))**0.5
\end{lstlisting}

\section{Evaluation}

Table 1 provides some reference scores as returned by the Python implementation given in
the previous section. Besides some new examples, we reproduce all the string 
comparison employed by Yang \textit{et al.} when presenting the MCWPA.

\begin{table}[htb]
\centering
\caption{Evaluation for MMCWPA}
\label{my-label}
\begin{tabular}{@{}lll@{}}
\toprule
$F_{X}$                           & $F_{Y}$                           & $SIM_{F_{(X, Y)}}$      \\
\midrule
\texttt{abc}                               & \texttt{def}                               & 0.0000 \\
\texttt{abcdef}                            & \texttt{abcdef}                            & 1.0000 \\
\texttt{Austria}                           & \texttt{Australia}                         & 0.6731 \\
\texttt{Python}                            & \texttt{python}                            & 0.8333 \\
\texttt{a123b}                             & \texttt{ab123}                             & 0.6982 \\
\texttt{129 Industry Park}                 & \texttt{129 Indisttry Park}                & 0.6101 \\
\texttt{abc de}                            & \texttt{abc k de}                          & 0.6388 \\
\texttt{de abc}                            & \texttt{de abc}                            & 1.0000 \\
\texttt{abc de}                            & \texttt{de abc}                            & 0.6236 \\
\texttt{Fu Hui}                            & \texttt{Mr Fu Hui}                         & 0.8000 \\
\texttt{Fu Hui}                            & \texttt{Fu Mr Hui}                         & 0.5962 \\
\texttt{abcdefgh ijklmnpo}                 & \texttt{abcdefgh ijklmnwo}                 & 0.8843 \\
\texttt{akabc axyz mo}                     & \texttt{aabc axyz muo}                     & 0.7768 \\
\texttt{abcdefagha}                        & \texttt{aijklamabc}                        & 0.3316 \\
\texttt{Gao Hua Ming}                      & \texttt{Gao Ming Hua}                      & 0.5892 \\
\texttt{zeng zeng}                         & \texttt{zeng hong}                         & 0.5983 \\
\bottomrule
\end{tabular}
\end{table}

\section{Conclusion}

This paper presented a new algorithm, Modified Moving Contracting Window
Pattern Algorithm (MMCWPA) for the calculation of field
similarity, strongly relying on previous work by Yang \textit{et al.} (which are in no
way associated with this work). As for MCWPA, theoretical analysis, concrete examples,
and experimental results indicate that MMCWPA improves the accuracy and efficiency
of the calculation of field similarity and should be considered alongside with
other field metrics, particularly when dealing with short strings representing
natural language.

\bibliographystyle{IEEEtran}

\end{document}